\documentclass[5p]{elsarticle}

\usepackage{graphicx}
\usepackage{mathptmx}   % use Times fonts if available
\usepackage{amsmath}	
\usepackage{graphics}
\usepackage[hypertexnames=false]{hyperref}
\usepackage{rotating}
\usepackage{amssymb}
\usepackage{color}

\begin{document}

\begin{frontmatter}

\title{Inverse population genetic problems with noise: \\ inferring extent and structure of haplotype blocks from point allele frequencies}

\author[1]{Oliver Keatinge Clay}

\ead{oliver.clay@gmail.com, oliver.clay@urosario.edu.co}

\affiliation[1]{organization={Translational Microbiology and Emerging Diseases (ITM/MICROS)}, addressline={\\ School of Medicine and Health Sciences, Universidad del Rosario}, city={Bogota}, postcode={111221}, country={Colombia}}

\newpageafter{author}

\begin{abstract}
A haplotype block, or simply a \emph{block}, is a chromosomal segment, DNA base sequence or string that occurs in only a few variants or types in the genomes of a population of interest, and that has an encapsulated or `private' frequency distribution of the string types that is not shared by neighbouring blocks or regions on the same chromosome. We consider two inverse problems of genetic interest: from just the frequencies of the symbol types (4 base types, possible single-base alleles) at each position (point, base/nucleotide) along the string, infer the location of the left and right boundaries of the block (block extent), and the number and relative frequencies of the string types occurring in the block (block structure). The large majority of variable positions in human and also other (e.g., fungal) genomes appear to be biallelic, i.e., the position allows only a choice between two possible symbols. The symbols can then be encoded as \texttt{0} (major) and \texttt{1} (minor), or as $\uparrow$ and $\downarrow$ as in Ising models, so the scenario reduces to problems on Boolean strings/bitstrings and Boolean matrices. The specifying of \emph{major allele frequencies} (MAF) as used in genetics fits naturally into this framework. A simple example from human chromosome 9 is presented.
\end{abstract}

\begin{keyword}
genotype-phenotype associations \sep haplotypes \sep standing variation \sep single-nucleotide polymorphisms \sep minor allele frequencies \\
\emph{Abbreviations}: $\,$ SNP: single-nucleotide polymorphism \sep MAF: minor allele frequency \sep kb: kilobase pair \sep  GA: genetic algorithm
\end{keyword}

\end{frontmatter}

\section{Introduction}
\label{prologue}

\noindent In this contribution, we present a simplified scenario or model developed while and after analyzing a number of block structures in DNA sequence segments of human chromosomes across two or more populations \cite{JACC1-JEG-OKC,GalloIJChy2020,Dingsdag2021}. In these blocks, we found that reported or \emph{de novo} observed genetic associations with phenotypic (physiological or clinical) traits were quite obviously delocalized, extending along the entire block. By contrast, \linebreak genotype-phenotype analyses are typically still being done in a base-by-base fashion, focusing on associations with single nucleotide polymorphisms (SNPs, i.e., variable base positions) and viewing covariation among such SNPs as held together by pairwise linkage disequilibrium between the SNPs. The problem with such an approach is that block structure then emerges only as a secondary property, and is recognized only indirectly via pairwise relations. A hallmark of this problem is that many studies still seek in vain to find a single `lead' SNP of a block that, in a given population, would have a maximally significant association with the trait of interest, in the hope of marking a position in the chromosomal sequence that has higher causal potential than all other SNPs in the same block. It is then sometimes concluded that fine mapping attempts `failed'. Instead, in the model we present here, we conclude that one must logically give identical credence to several SNPs because they are completely equivalent (unless one can, later, differentiate them on the basis of detailed, `external' experimental knowledge of the molecular biology of the specific DNA sequence of the block, which is very often not available).

Our stance puts structure first, and then quantitatively predicts from the block structure how the frequencies of the states (alleles) at the different point positions (SNPs) must be related to each other. Here, we take a phenomenological viewpoint, and do not focus on details of how structures arose, e.g., from the merging of historic populations that had long evolved independently, with little if any surviving evidence for viable recombination (crossover, chiasma) events having occurred in a block's interior \cite{BartonHaplotypes2023,SabetiLander2024}. The resulting structure is sometimes considered as an ancestral population structure or admixture structure, and the main genetic variation it represents could be considered to be standing variation. A current trend has been to identify and effectively remove population structure or stratification before looking for genotype-phenotype associations, for example by principal component analysis and the discarding or ignoring of the first 3-20 components, as if they were a nuisance factor \cite{Uffelmann2021}. Although this aim may be useful where a sample from a population was only recently stratified, the presence of natural and ancestrally derived standing variation (which could allow selection to act more rapidly) would seem valuable to retain for revealing associations that efficiently inform on molecular mechanisms contributing to phenotypic traits.

The next section presents an introductory toy example, the following section shows a real-world scenario, and a final section focuses on questions that can be partly abstracted from the biological contexts.

\section{A simple simulation}

\noindent We first present and discuss a simple toy simulation exercise, essentially from the 1980's, i.e., well before publication of the first haplotype blocks that were found in complete human chromosome sequences at the turn of the century.\footnote{Haplotype blocks, observed along human chromosomes, are conceptually similar to blocks that were observed much earlier in simulations \cite{HollandBook1995GAs,GoldbergBook1989GAs} and named \emph{building blocks}.}

We imagine a population of 100 haploid digital organisms, each with a genome consisting of a single chromosome of length five bases. At each of the five base positions or sites, we allow a biallelic polymorphism (i.e., a single nucleotide polymorphism, or SNP, with two possible alleles or states). At each of those 5 sites we code the two possible alleles as 0 and 1. We observe the population of the 100 organisms (i.e., chromosomes) during a time interval within which the relative fitnesses of the $2^5$ imaginable chromosome variants (types) do not change. We use a simple model, fitness rule and evolutionary simulation algorithm from one of the first books from the pioneer epoch of genetic algorithms (GA) \cite{GoldbergBook1989GAs}, and its implementation in C (SGA-C\footnote{https://www.cs.cmu.edu/afs/cs/project/ai-repository/ai/areas/genetic/ga/ \\
systems/sga/sga\_c/0.html}).
The model is for discrete generations that do not overlap. The 32 possible 5-nt genomes (chromosomes, bitstrings) can be interpreted, reading from right to left (i.e., `backwards' compared to the usual direction), as the binary representations of the first 32 integers $n$ starting at 0, i.e.: 0,1,\ldots 31$= n_{\mathrm{max}}$. In the model, the relative fitness \cite{GoldbergBook1989GAs}, or viability \cite{EdwardsFoundations}, of each genome is $(n/n_{\mathrm{max}})^{10}$; the book gives some justification for this choice. Thus, the highest fitness is assigned to the haplotype 11111 and the lowest fitness to the haplotype 00000. Furthermore, the fitness model sets up a strong selection-pressure gradient, increasing from left (least significant bit) to right (most significant bit): substitutions of nucleotides (bits) at the right end will have dramatic effects on fitness, while the effects of substitutions at the left end will be far smaller.\footnote{Indeed, imagine a single chromosome of the population which mutates from 11111 via a single point substitution. This substitution or mutation could occur at any of the five positions $i$, which we number 0,1 2,3,4 from left to right. The loss in fitness incurred by such a mutation is then $1 - ((31-2^i)/31)^{10}$, i.e., at each of the 5 positions the mutation would incur a fitness loss of 0.280, 0.487, 0.748, 0.949 and 0.999 respectively, out of a maximal fitness loss of 1.} The algorithm takes care of updating or `renormalizing' at each (non-overlapping) generation in such a way that the total population size of the following generation is again 100 (\cite{GoldbergBook1989GAs}; see also \cite{EdwardsFoundations}).

Our initial population is now generated in a completely random fashion (we choose the seed 0.4141); thereafter, the haplotype frequency ratios are updated in such a way that a fitter variant in the haplotype pool is typically `rewarded' in the next generation by a higher viable replication rate, i.e., an increased relative representation in the population \cite{GoldbergBook1989GAs,EdwardsFoundations}. For example, if the strings for the three fittest integers 31, 30 and 29 (haplotypes 11111, 01111, 10111; fitnesses 1, 0.720436, 0.513290) all happen to be present in the initial, randomly generated primordial pool, then their frequencies are likely to rise in subsequent generations. 

Ultimately, our static-fitnesses model will push the fittest to take over completely and become the only haplotype at the end (monomorphism, fixation). Here, we are interested in \emph{the time window during which the co-existence/polymorphism of just a few haplotype variants is maintained}, i.e., before fixation happens. The theory that describes events and evolution during this time window we will call \emph{few-haplotype theory}. 

We again point out that throughout the simulations, the relative fitnesses of the variants are assumed not to change with time. In real situations, occurring in real and possibly fluctuating environmental conditions, fitnesses sometimes change (and fitness rankings invert) before fixation can be reached. In such shifting conditions, the population may not reach a final showdown among two or few survivor haplotypes that one might otherwise anticipate. Indeed, where fitnesses keep changing, fixation may not occur until much later, if at all.

We now turn off the simulator's mutation and crossover generators, and start the run.

\begin{figure}
\hspace*{1.3cm}\includegraphics[width=0.35\textwidth]{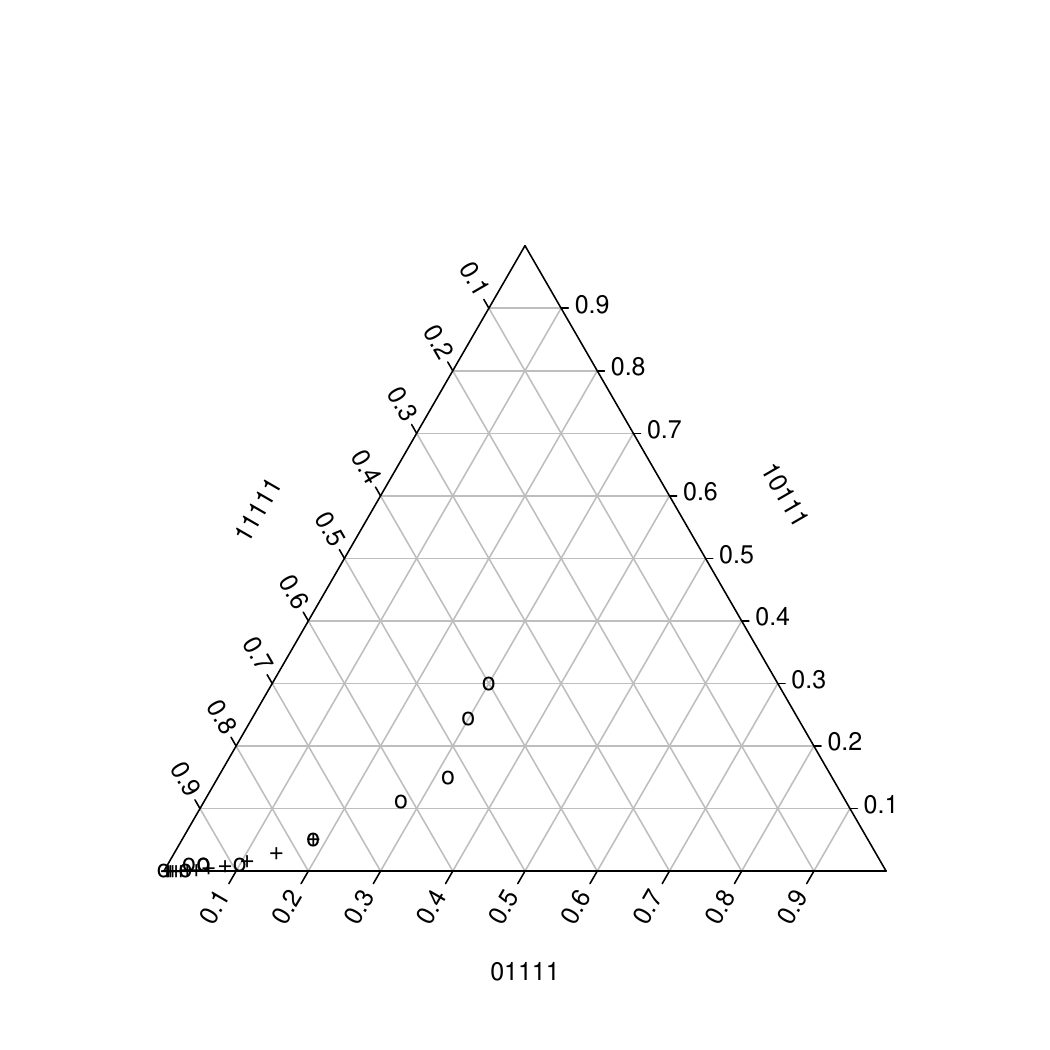}
\caption{Relative frequencies of the three dominating haplotypes during the population's trajectory or journey toward fixation (bottom left vertex of trangle). Circles and crosses represent values from the simulation ($t \ge 0$) and the map ($t \ge 4$), respectively. This ternary plot was produced using the R package \texttt{plotrix}.}
\label{fig:zero}
\end{figure}

The following display shows the results of the simulation. The first column shows the distribution of chromosome types (haplotypes) in the initial, randomly generated population at time $t=0$. The remaining columns show the haplotype distributions in the 10 following generations, which in this case suffice for the fittest haplotype to go to fixation. At each generation, the haplotypes are shown together with the counts (or, in this simple example, frequencies, expressed as percentages) of the different 5-nt haplotypes (types), sorted by descending count. The three haplotypes we can track from the beginning until and including $t=8$ are shown in bold. These three fittest haplotypes have the highest of all frequencies already after the first generation; after an additional generation they already account for 94\% of the chromosomes.

\bigskip

\scriptsize
\hspace*{-0.65cm} \begin{tabular}{|lr|lr|lr|lr|lr|}
$t=0$: & & $t=1$: & & $t=2$: & & $t=3$: & & $t=7$: & \\
 & & & & & & & & & \\
00010 & 6 & \textbf{11111} & 32 & \textbf{11111} & 50 & \textbf{11111} & 61 & \textbf{11111} & 94 \\
10110 & 6 & \textbf{01111} & 21 & \textbf{01111} & 30 & \textbf{01111} & 27 & \textbf{01111} & 5 \\ 
00000 & 5 & \textbf{10111} & 17 & \textbf{10111} & 14 & \textbf{10111} & 11 & \textbf{10111} & 1 \\
00011 & 5 & 11011 & 12 & 00111 & 3 & 00111 & 1 & & \\
11000 & 5 & 00111 & 11 & 11011 & 2 & & & $t=8$: & \\
01000 & 4 & 01011 & 4 & 01011 & 1 &  $t=4$: & &   & \\      
01010 & 4 & 00011 & 2 & & & & & \textbf{11111} & 96 \\      
01011 & 4 & 10110 & 1 & & & \textbf{11111} &  76 & \textbf{01111} & 3 \\
01100 & 4 & & & & & \textbf{01111} & 18 & \textbf{10111} & 1 \\
11010 & 4 & & & & & \textbf{10111} & 5 & & \\    			
11011 &	4 & & & & & 00111 &	1 & $t=9$: & \\
11110 &	4 & & & & & & &   & \\
\textbf{11111} & 4 & & & & &  $t=5$: & & \textbf{11111}  & 97 \\
00110 &	3 & & & & & & & \textbf{01111} & 3 \\
00111 &	3 & & & & & \textbf{11111} & 89 & & \\
01101 &	3 & & & & & \textbf{01111} & 10 &  $t=10$: & \\
01110 &	3 & & & & & \textbf{10111} & 1 & & \\
\textbf{01111} & 3 & & & & & & & \textbf{11111} & 100 \\
10001 &	3 & & & & &  $t=6$: & & & \\
10010 &	3 & & & & & & & & \\
10100 &	3 & & & & & \textbf{11111} &  94 & & \\
\textbf{10111} & 3 & & & & & \textbf{01111} & 5 & & \\
00100 &	2 & & & & & \textbf{10111} & 1 & & \\
00101 &	2 & & & & & & & & \\
01001 &	2 & & & & & & & & \\
10000 &	2 & & & & & & & & \\
10011 &	2 & & & & & & & & \\
11001 &	2 & & & & & & & & \\
00001 &	1 & & & & & & & & \\
10101 &	1 & & & & & & & & \\
\end{tabular}
\normalsize

\bigskip

The frequencies expected at each generation can also be calculated analytically, instead of simulating them; if we consider just the three bold haplotypes and label them 0, 1 and 2 respectively, the well-known maps for this haploid case are \cite{EdwardsFoundations}

\vspace*{-0.3cm}\begin{eqnarray*}
f_i & \mapsto & f_i \, ' = w_i f_i/w \;\;\;\;\;\; (i \in \{0,1,2\}) \; , \\
\end{eqnarray*} 

\vspace*{-0.5cm} \noindent where the $w_i$ are the relative fitnesses or viabilities of the three haplotypes and $w$ is a normalizing constant called the (weighted) average fitness or mean viability \cite{EdwardsFoundations}, which ensures that the frequencies will add up to 1 in each generation; in our case it is also the proportion of haplotypes that survives from birth to age of replication \cite[p.\ 5]{EdwardsFoundations}. Here, and in the sequel, $f_i$ denotes the frequency of the $i$-th haplotype, with numbering beginning at 0 (for a reason that will be explained later).

We now compare how the map will project the expected frequencies from the 4th generation ($t=4$) into the future. Given our previous fitness allocation, $(n/n_{\mathrm{max}})^{10}$, the haplotypes numbered $n =$ 31, 30, 29 had fitnesses $w_1=1$, $w_2=0.720436$, $w_2=0.51329$. At $t=4$ we had frequency ratios of 76 : 18 : 5, so for the next generation the map would predict about  83 : 14 : 3. The agreement with the simulation's result (see table) of  89 : 10 : 1 is satisfactory, given the inherent stochasticity of a small population of size 100. The ternary plot in Figure 1 shows the full trajectory (from $t=0$) of the GA simulation as circles and, superposed, the map's projection of the expected values as crosses (from $t=4$ onwards).

Clearly, from $t=3$ onwards we would no longer need to keep track of the four rightmost bits, because they are always 1: the sites they represent are no longer polymorphic but \emph{monomorphic}, i.e., they do not change. Also in the sequel, we will not consider the monomorphic sites of a haplotype.

Finally, we recode with the \emph{minor coding} at each position: we code the minor allele at each position, i.e., the `mutant' or exception, as 1 and the other, more frequent allele or `wildtype', as 0. In our toy exercise, it so happens that all 1's will now need to be recoded as 0's and vice versa.

Economizing and then recoding in this way, we can display the frequencies of the three main haplotypes and their frequencies as follows:

\begin{tabular}{cc|c}
$\,$ \\
\texttt{0} &  \texttt{0} & $f_0$ \\[3pt]
\texttt{1} &  \texttt{0} & $f_1$ \\
\texttt{0} &  \texttt{1} & $f_2$ \\
\hline
$g_1$ &  $g_2$ &  \\
$\,$
\end{tabular}

\noindent Here, we again denote the frequencies of the haplotypes by $f_i$ ($i \in \{0,1,2\}$), but the frequencies of the SNP minor alleles by $g_j$ ($j \in \{1,2\}$), i.e., numbering from 1 instead of 0. 

There is thus a duality between the frequencies/counts of the haplotypes (e.g., those shown in Table 1 for a given generation $t$) and the frequencies of the two possible nucleotide (base) types at the individual positions along those haplotypes, i.e., the `vertical' counts of 1's and 0's in individual positions of the haplotypes.

An inverse problem is to use the frequencies of the single positions' (sites', bases', nucleotides') variants, i.,e., the individual positions' allele frequencies, to infer the haplotypes responsible for them, and the frequencies of those haplotypes. There is practical relevance and utility of such inverse reasoning, especially when one seeks to know where a phenotypic trait of interest is associated with genetic variants.

The 0th haplotype is just the null row, so we can visually set it apart (or ignore it) and focus on the more informative 2 $\times$ 2 `core' matrix below it, which in this toy example is the identity matrix, its own inverse, so the inverse problem is trivially solved.

Generally (i.e., not just for this toy example), if we use the minor coding and denote the elements of the full matrix by $u_{ij}$, then we have the (forward) conversion relation

\begin{displaymath}
g_j \;  = \; \sum_{i=0}^2 u_{ij} f_i \;  = \;  \sum_{i=1}^2 u_{ij} f_i \; ,\\
\end{displaymath}

\noindent where the second identity holds if the coding has been arranged so that the 0th (most frequent) haplotype is the null row.

Here a clarification is needed. If a block has more than two haplotypes, then the allele at a given position in the major (most frequent) haplotype need not always be the major allele coded at that position only, which might prevent the major haplotype from being the null row or null string. Assuming as usual that all sites are biallelic, we propose the following

\smallskip

\noindent \textbf{Definition}. A \emph{consistent minor coding} of a block is a minor coding in which the major allele at each position in the block is also the position's allele in the major haplotype, allowing consistent coding of that allele as \texttt{0}.

\smallskip

That not every block allows a consistent minor coding is shown by a simple

\smallskip

\noindent \textbf{Counterexample}. Consider a block having only three haplotypes, with frequencies 0.4, 0.3 and 0.3, in which the haplotypes have at a given position (site) the alleles \texttt{A}, \texttt{C} and \texttt{C}, respectively. If one considers the position in isolation, one would need to code \texttt{C} as 0 because its allele frequency at that site is 0.6, so the major haplotype would have a 1 at that position, i.e., the major haplotype would not be the null row.

\smallskip

However, in blocks containing only two kinds of haplotype (so-called `yin-yang' blocks \cite{gephyrinYinYang}) every minor coding is consistent, because allele frequencies in such blocks do not lose information or `resolution' by aggregating haplotype frequencies. 

\section{A real-world example from chromosome 9}

\noindent We are now ready to consider a `noisy' real-world example, from cytogenetic band 9p21.3 on chromosome 9, which also will motivate the final sections. Unlike the haploid organisms of the toy example, human genomes are diploid, i.e., they have two slightly different copies of each chromosome. However, aligning the sequences, two per person, allows the total number of each of the two alleles at each site (the SNP alleles) to be counted exactly.\footnote{At this time, it is not yet easy or inexpensive to obtain or reconstruct precise end-to-end chromosome sequences of each of the two slightly different copies of a given chromosome of a person, i.e., to get the precise sequences of the person's two separate haplotypes \cite{Eichler2021}. This might be a practical or economic motivation for the inverse problems mentioned, although haplotype reconstructions by 1000 Genomes \cite{KGgenomes} using trios and the Beagle program\cite{Beagle-v4-2013} have been quite good.} We will not consider here the very few non-biallelic SNPs of the human genome; indeed, in all human populations and their individuals sampled in the 1000 Genomes project, over 99.6\% of the SNPs were biallelic \cite[Supplementary Table 2]{KGgenomes}.\footnote{At first sight, such statistics might suggest that many-allele theory (see, e.g., \cite[Chapters 4-5]{EdwardsFoundations}) would have limited utility. However, precisely the view sketched here suggests utility at a much larger scale than that of SNPs, namely that of entire haplotype blocks. Indeed, founder concepts of genetics were established well before the structure of DNA or the genetic code were known, and early concepts such as \emph{allele} and \emph{locus} were inherently and necessarily scale-free. For example, three-allele theory can be profitably applied to a block having three haplotype variants or classes, i.e., three \emph{haplotype alleles} or allele classes (see Figure 1); for diploidy, the SNP genotypes correspond at the larger scale to \emph{diplotypes} \cite[Supplementary Material]{GalloIJChy2020}.}

We will also restrict attention to \emph{common} SNPs, defined here as those having a \emph{minor allele frequency} (MAF) greater than 0.03, i.e., a presence of the minor allele in at least 3\% of the 5008 haplotypes of the 1000 Genomes study.

The following diagram, of a fictitious, very short block (modified from \cite{DalyLanderHapblocks}), sampled here for 5 persons, recapitulates some of the names and conventions introduced so far. For the coding, we use the minor coding \emph{for haplotypes} to guarantee the null string for the most frequent haplotype in this block, which allows a consistent minor coding. The final table keeps one instance of each of the two column types, or \emph{SNP classes}:

\bigskip

\newlength{\oldtabcolsep}
\setlength{\oldtabcolsep}{\tabcolsep}
\setlength{\tabcolsep}{0pt}

\hspace*{-0.3cm}\begin{tabular}{lclclll}
1a	& \texttt{CG\underline{G}A\underline{A}CGA} & & \texttt{00\underline{0}0\underline{0}000} & & & \\ 	
1b	& \texttt{GA\underline{C}T\underline{G}TCG} & & \texttt{00\underline{0}0\underline{0}000}  & & & \\ 
2a	& \texttt{CG\underline{C}A\underline{A}CGA} & & \texttt{00\underline{0}0\underline{0}000}  & & & \\ 
2b	& \texttt{CG\underline{G}A\underline{A}CGA} & & \texttt{00\underline{0}0\underline{0}000}  & & \texttt{0} \texttt{0} & $\;\;\;$ 0.6 \\ 
3a	& \texttt{CG\underline{G}A\underline{A}CGA} & $\;$ $\mapsto$ & \texttt{00\underline{0}0\underline{0}000}  &  $\;\;\;\;\;\;$ $\mapsto$ & \texttt{1} \texttt{1} & $\;\;\;$ 0.3 \\
3b	& \texttt{CG\underline{G}A\underline{A}CGA} & \emph{sort},$\;\;\;\;$ & \texttt{00\underline{0}0\underline{0}000}  & $\;\;\;\;$ \emph{table} $\;\;\;\;$ & \texttt{1} \texttt{0} &  $\;\;\;$ 0.1 \\
4a	& \texttt{GA\underline{C}T\underline{G}TCG} & \emph{code} $\;\;\;\;$ & \texttt{11\underline{1}1\underline{1}111}  & & & \\ 
4b	& \texttt{CG\underline{G}A\underline{A}CGA} & & \texttt{11\underline{1}1\underline{1}111}  & &  & \\
5a	& \texttt{CG\underline{G}A\underline{A}CGA} & & \texttt{11\underline{1}1\underline{1}111}  & &  & \\	
5b	& \texttt{GA\underline{C}T\underline{G}TCG} & & \texttt{00\underline{1}0\underline{0}000}  & & & \\
 &  $\;\;\;$\begin{sideways}$\longrightarrow$\end{sideways} & & $\;\;\;$\begin{sideways}$\longrightarrow$\end{sideways} & & & \\
 & SNP, nucleotide, & & SNP alleles & & & \\
 & site, base, point, & & coded \texttt{0},\texttt{1} & & & \\ 
  & position, with & & & & & \\ 
  & SNP alleles \texttt{A},\texttt{G} & & & & & \\ 
\end{tabular}

\setlength{\tabcolsep}{\oldtabcolsep}

\bigskip

One can quite easily generate, from the 1000 Genomes phase 3 database, a \emph{MAF plot} of the common biallelic SNPs in a region, such as the 200 kb region of chromosome 9 (9p21.3) shown in the following 5-tier MAF plot, for the five (super-) populations of the 1000 Genomes study:

\begin{center}
\includegraphics[width=0.55\textwidth]{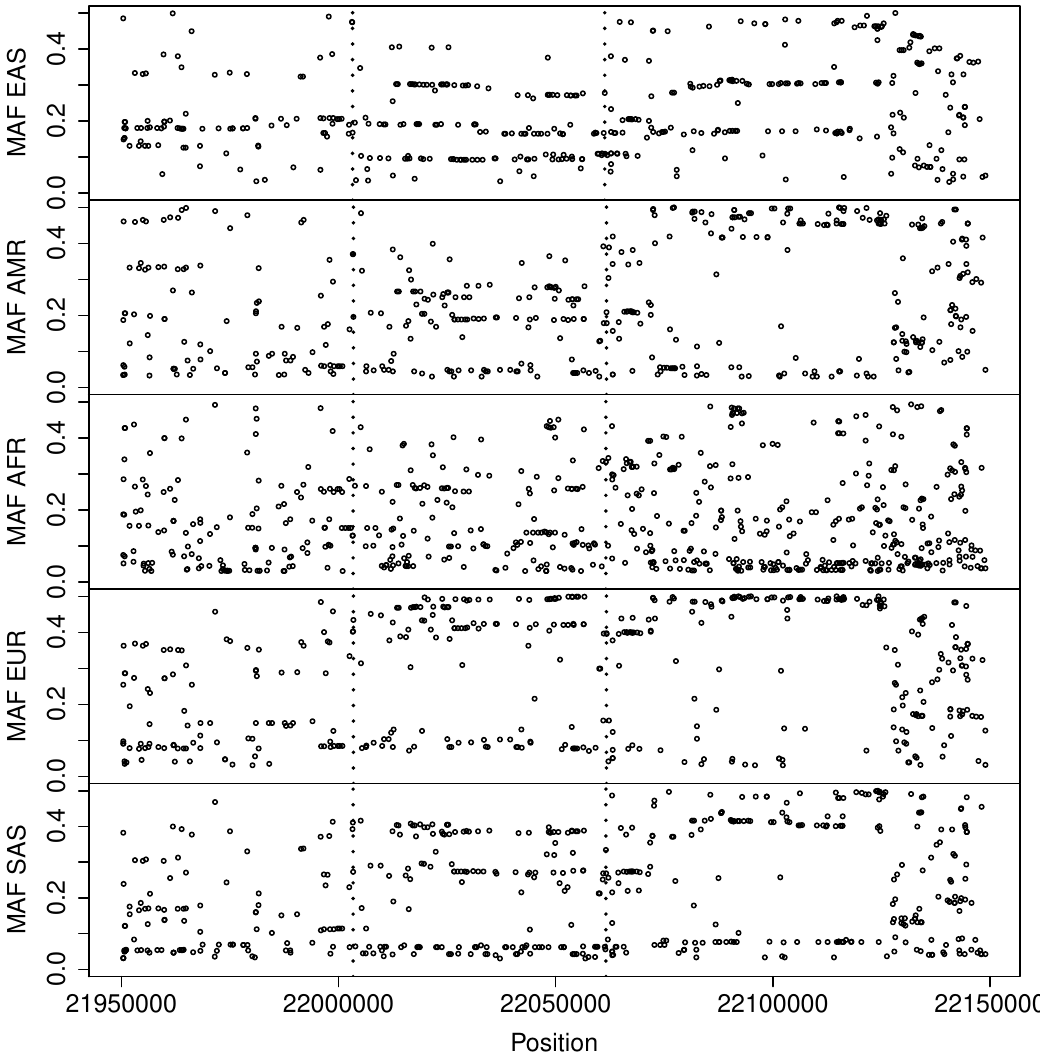}
\end{center}

\noindent Here, we again ignore monomorphic sites and rare SNPs. Each point represents a single biallelic common SNP that is polymorphic for the (super)population shown, at the chromosomal position shown (human genome release hg18). The horizontal `stripes' or rows consist of repeated occurrences of essentially the same few MAFs persisting across a wide block, reflecting the underlying haplotype frequency distribution of the block. One can see here graphically that it would be attractive to solve the obvious inverse problems, in the chromosomal regions and populations where a block structure is visible: from the persistent MAFs in a candidate block, determine the number and frequencies of the (main) haplotype types or haplotype classes\footnote{What one actually observes in many if not most real data sets is that haplotypes of a block will cluster naturally into very few haplotype \emph{classes}. Within each of those few classes there is usually one most frequent sequence, or \emph{master sequence} or \emph{master haplotype}, that dominates in frequency. Within each haplotype class there is often remarkably little MAF variation, compared to the sometimes dramatic MAF contrasts observed among the haplotype classes of the block.}, and estimate the locations of the block's boundaries. The region has long been of interest ever since the first major WTCCC genome-wide association study in 2007, because of its multiply confirmed associations with cardiovascular disease and risk factors \cite{WTCCC2007,GalloIJChy2020}. A detailed analysis of the haplotype block at the center of the region (between the two dotted vertical lines), in AMR (populations from the Americas having a Native American ancestry component, $n=$694 individual haplotypes) and EUR (European ancestry, $n=$1086 haplotypes), is given in \cite{GalloIJChy2020} and involves, as in many real problems, the replacing of haplotype types by (often easily identified) haplotype \emph{classes}. A caricature of the structure of the central block of this pleiotropic locus in the AMR population is given by the following simple frequency table:

\smallskip

\begin{center}
\begin{tabular}{ccc|r}
	0 & 0 & 0 & .70 \\[3pt]
	1 & 1 & 0 & .24 \\
	1 & 0 & 1 & .06 \\
	\hline
	.30 & .24 & .06 &\\
\end{tabular}
\end{center}

\smallskip

Wherever different chromosomal positions (SNPs) within a block have identical MAFs, it is typically because there is (near) identity of their full Boolean SNP-allele state vectors (columns) showing the allele of each individual in a sample or cohort. Since SNP-wise $p$-values for associations with traits are calculated using precisely that information, $p$ values at the positions will also be equal or similar, unless data are differentially missing or imputed. As a result, there will be not just one robust `lead' SNP of the block, but a set of equivalent lead SNPs. The problems go deeper, however, if the association is not with individual SNPs but with the entire block. Several artefacts are conceivable in such a situation. (1) If one insists on attributing associations only to individual SNPs, and the block is composed of three or more types (or classes) of haplotype, one can fail to recognize a risk or protector haplotype, and instead get only aggregate-based results at individual SNPs that do not faithfully reveal the true effects (size and/or direction), i.e., the underlying trait contrasts among haplotypes. (2) SNPs near the fringe of a block could be fortuitously given an advantage (lower $p$ value for an association test) and named lead SNP of the block merely because of haplotype length heterogeneity (`ragged boundaries')\footnote{A simplistic example, for a dichotomous (case/control) trait: At a single-base position in the ragged boundary zone at the fringe of a block of a population's sample, one of 350 control haplotypes from those individuals has a 1 although elsewhere in the block it has a 0, such that the fringe SNP's $2 \times 2$ table $\chi^2$-test $p$-value is (in R) \texttt{chisq.test(matrix(c(350,349,100,201),nr=2))\$p.value} = $1.28 \times 10^6$, lower (i.e., more significant) than the $p$-value $1.71 \times 10^6$ for \texttt{c(350,350,350,100,200)} inside the block.}; the previously identified lead SNPs rs1333049 and rs1333050 for the 9p21.3 risk locus \cite{WTCCC2007}, located at the top and halfway down the rightmost cascade of points in the MAF plots above, may or may not be a case in point. (3) Two populations having the same three haplotypes but with very different frequencies may, because of the aggregating of counts in SNP-based testing, show an effect in one population but not in the other because a pair of haplotypes there had mutually canceling deviations of trait values.

We end this section with an initial summary of such situations, from the paper that first reported the widespread presence of haplotype blocks in the human genome shortly after its first sequencing in 2001  \cite{DalyLanderHapblocks}\footnote{One may wonder why this advice has hardly been taken into consideration during the constructing of association genetics pipelines and protocols since it was written (see however \cite{SabetiLander2024}). One reason may be that the initial discovery of haplotype blocks soon led to unrealistic hopes that an exhaustive, gapless segmentation of `the' human genome into haplotype blocks would be possible, valid across many world populations, and when this aim was shown to be unfeasible the disappointment dissuaded further research along such lines. Nevertheless, trait-associated loci frequently do have a marked block structure that is visible in several world populations.}:

\begin{quotation}
\noindent ``... the haplotype structure provides a crisp approach for testing the association of genomic segments with disease. By contrast, disease association studies traditionally involve testing individual SNPs in and around a gene. This approach is statistically weak and has no clear endpoint: true associations may be missed because of the incomplete information provided by individual SNPs; negative results do not rule out association involving other nearby SNPs.''
\end{quotation}

\section{Haplotype frequencies and SNP state vectors}

\noindent In some blocks, the haplotype structure in extant human populations might simply not offer enough of the right kind of natural variation, in the sense of generating enough different SNP classes, for the inverse problems to be solvable. In other words, the rank of the core matrix (which was 2, i.e., maximal, in the unit matrix of the first toy example we considered above) may be too low.

The major/minor allocation (e.g., using the minor coding) is a relatively robust way of bypassing allele labeling problems\footnote{Integrating allele specifications from different sources can be fragile: a base pair viewed on one strand (reading direction) is reported as \texttt{A}, but another researcher choosing as reference the opposite strand may instead report its complement (\texttt{T}), without mentioning which sense was chosen, so results for an A/T SNP may be difficult to interpret or compare. This problem is sometimes seen when integrating results from the NHGRI-EBI GWAS database.}, but its elegance is compromised when MAFs fluctuate around 50\%. A MAF plot folds, by definition, the interval [0,1] back onto itself, i.e., maps it onto [0,0.5]; this requires caution when completing frequency tables, since  a horizontal stripe in the MAF plots above that is located at height 0.3 could represent either a SNP frequency $g_j$ = 0.3 in that table, or its complement 1 $-$ $g_j$ = 0.7. Such ambiguities can often be resolved because the haplotype frequencies $f_i$ must add up to 1.

If one uses the minor coding \emph{of haplotypes}, in which the most frequent haplotype is always represented by the null string \texttt{00....0}, it is easy to generate and list the different SNP classes, i.e., types of SNP state vectors (columns). For 2, 3, 4, 5, 6, and 7 haplotype classes there can be 1, 3, 7, 15, 31, and 63 possible minor allele frequencies (MAFs), respectively. A biallelic SNP class with its allele frequencies corresponds to a partition of a set of haplotypes into two disjoint subsets. Each such partition corresponds to a possible column (state vector) of the basic 0/1 incidence matrix, and to a possible minor SNP-allele frequency (MAF). The number of partitions of any set rises rapidly with the number of elements of the set. The number of ways in which one can partition a set of $n$ elements into two non-empty subsets is given by the corresponding Stirling number of the second kind, or Stirling partition number, $S(n,2)$, which is equal to 2$^{n-1}$ $-$ 1. However, in the usual presence of some minor fluctuations (noise), the distinguishing of the different haplotype types/classes becomes practically intractable if there are more than 5 types or classes, and often well before that count of 5, if two or more classes have very similar frequencies.

The different MAFs or SNP classes will correspond to the (typically different) SNP-by-SNP $p$-values found if one screens for associations with a trait in that haplotype block. Consequently, where there is a block structure, the so-called Manhattan plots, of $-\log p$ versus chromosomal position, that often accompany association discovery reports will tend to have a horizontally striped pattern that corresponds largely to that of the MAF plot of the same region in the same population.

The tables showing possible SNP classes and their frequencies for 3, 4 and 5 haplotype types, or master haplotypes representing haplotype classes, are as follows:

\small

\begin{center}
\begin{tabular}{ccc|r}
0 & 0 & 0 & $f_0$ \\[3pt]
1 & 1 & 0 & $f_1$ \\
1 & 0 & 1 & $f_2$ \\
\hline
\!\!\!\!$1-f_0$\!\!\! & $f_1$ & $f_2$ & \\
\end{tabular}
\end{center}

\begin{center}
\begin{tabular}{ccccccc|r}
0 & 0 & 0 & 0 & 0 & 0 & 0 & $f_0$ \\[3pt]
1 & 1 & 0 & 0 & 1 & 1 & 0 & $f_1$ \\
1 & 0 & 1 & 0 & 1 & 0 & 1 & $f_2$ \\
1 & 0 & 0 & 1 & 0 & 1 & 1 & $f_3$ \\
\hline
\begin{sideways}$1-f_0\,\,$\end{sideways} & 
\begin{sideways}$f_1\,\,$\end{sideways} & 
\begin{sideways}$f_2\,\,$\end{sideways} & 
\begin{sideways}$f_3\,\,$\end{sideways} & 
\begin{sideways}$f_1+f_2\,\,$\end{sideways} & 
\begin{sideways}$f_1+f_3\,\,$\end{sideways} & 
\begin{sideways}$f_2+f_3\,\,$\end{sideways} & 
\end{tabular}
\end{center}

%\newlength{\oldtabcolsep}
\setlength{\oldtabcolsep}{\tabcolsep}
\setlength{\tabcolsep}{2pt}

%\scriptsize
\begin{center}
\begin{tabular}{ccccccccccccccc|r}
0 & 0 & 0 & 0 & 0 & 0 & 0 & 0 & 0 & 0 & 0 & 0 & 0 & 0 & 0 & $\;$ $f_0$ \\[3pt]
1 & 1 & 0 & 0 & 0 & 1 & 1 & 1 & 0 & 0 & 0 & 1 & 1 & 1 & 0 & $\;$ $f_1$ \\
1 & 0 & 1 & 0 & 0 & 1 & 0 & 0 & 1 & 1 & 0 & 1 & 1 & 0 & 1 & $\;$ $f_2$ \\
1 & 0 & 0 & 1 & 0 & 0 & 1 & 0 & 1 & 0 & 1 & 1 & 0 & 1 & 1 & $\;$ $f_3$ \\
1 & 0 & 0 & 0 & 1 & 0 & 0 & 1 & 0 & 1 & 1 & 0 & 1 & 1 & 1 & $\;$ $f_4$ \\
\hline
\begin{sideways}$1-f_0\,\,$\end{sideways} & 
\begin{sideways}$f_1\,\,$\end{sideways} & 
\begin{sideways}$f_2\,\,$\end{sideways} & 
\begin{sideways}$f_3\,\,$\end{sideways} & 
\begin{sideways}$f_4\,\,$\end{sideways} & 
\begin{sideways}$f_1+f_2\,\,$\end{sideways} & 
\begin{sideways}$f_1+f_3\,\,$\end{sideways} & 
\begin{sideways}$f_1+f_4\,\,$\end{sideways} & 
\begin{sideways}$f_2+f_3\,\,$\end{sideways} & 
\begin{sideways}$f_2+f_4\,\,$\end{sideways} & 
\begin{sideways}$f_3+f_4\,\,$\end{sideways} & 
\begin{sideways}$f_1+f_2+f_3\,\,$\end{sideways} & 
\begin{sideways}$f_1+f_2+f_4\,\,$\end{sideways} & 
\begin{sideways}$f_1+f_3+f_4\,\,$\end{sideways} & 
\begin{sideways}$f_2+f_3+f_4\,\,$\end{sideways} & 
\end{tabular}
\end{center}

\setlength{\tabcolsep}{\oldtabcolsep}

% $1-(f_0+f_4)\,\,$ is $f_1+f_2+f_3\,\,$
% $1-(f_0+f_3)\,\,$ is $f_1+f_2+f_4\,\,$ 
% $1-(f_0+f_2)\,\,$ is $f_1+f_3+f_4\,\,$  
% $1-(f_0+f_1)\,\,$ is $f_2+f_3+f_4\,\,$   

\normalsize

We end this section with a simple (contrived) example that uses the first of these three tables (for $n=3$ haplotypes) to illustrate the kind of ramifications one may expect if one ignores haplotype structure when screening for associations, and instead performs only traditional SNP-by-SNP testing.

\smallskip

\noindent \textbf{Example}. Consider a block having 3 haplotypes, with frequencies 0.70, 0.15 and 0.15. We assume, as is sometimes done in bulk genome-wide testing \cite[Table 1]{Ehret2016}, that a continuous trait of interest is allele-additive (i.e., not dominant, recessive or overdominant); we can then attribute a formal (mean) trait value not only to genotypes (or diplotypes) but also to individual alleles (or haplotypes). We assume that the association is inherently with a whole block and not just its individual SNPs, and seek to illustrate how far the set of 3 `effect' sizes, for the contrasts in the block's 3 SNP classes obtained individually, could deviate from the set of 3 pairwise effect sizes that actually characterize the 3 haplotypes. If the trait values for the 3 haplotypes are 0, 1 and $-$1, then the apparent (aggregate) contrast sizes of the SNPs, routinely calculated for the 3 columns of the $n=$3 table above, are 0 for SNPs of one SNP class and 20/17=2/1.7 for SNPs of the other two classes\footnote{If the (mean) trait values of the haplotypes with frequencies $f_0$, $f_1$ and $f_2$ are $m_0$, $m_1$ and $m_2$, then the three aggregate contrasts are $|m_i - (m_j f_j + m_k f_k)/(f_j+f_k)|$ for $i,j,k$ in $(012)$.}, whereas the pairwise contrast sizes among the haplotypes are 1, 1 and 2. The biggest (true) effect size for the entire haplotype block is therefore 70\% higher than the biggest (apparent) effect size that any of its individual SNPs report.

\smallskip

\section{Conclusion}

\begin{quotation}
\noindent ``Thus, if only sequences of a certain distinguishable group are used to send signals, then, despite the noise, the signals sent can be guessed with overwhelming probability.'' %
\begin{flushright}
A.\ I.\ Khinchin, \emph{Mathematical Foundations\\
of Information Theory}, Dover, 1957, p.\ 93.
\end{flushright}
\end{quotation}

\noindent This contribution presents, in a simplified form, a view of genetic variation and covariation with traits that is likely to be relevant in some regions of human and other genomes, in particular in some that are associated or potentially associated with phenotypic traits of physiological or medical importance \cite{JACC1-JEG-OKC,GalloIJChy2020,Dingsdag2021,VelezMM2023}. In the author's opinion the relatively simple, non-standard quantitative approach sketched in this research element has potential if pursued or developed further. 

In such contexts, the signals we must guess (in the quoted sentence of Khinchin) are transmitted via repetitive or `redundant' use of SNP classes and frequencies, and could include discovery elements revealing details of molecular mechanisms or disease etiologies.\footnote{As the quote from Khinchin suggests, a metaphor related to the view presented here might be that of a genetic channel \cite{MargalefBook}, via which information (e.g., for conveying molecular or regulatory strategies) is transmitted to the future via genetic change, and also to us for understanding and then modulating or correcting. A block could then be seen as a part of this effort, setting a course by using its possible SNP classes, involving dynamic or rapid selection effecting change in response to temporally and geographically local conditions and cues.} Such repetitive use across an entire block, often visible in MAF plots, gives us confidence that there is local coherence of the message, and that as observers we are on the right track.

\section*{Acknowledgments} 

\noindent I thank the co-authors of refs. \citenum{JACC1-JEG-OKC,GalloIJChy2020,Dingsdag2021} and \citenum{VelezMM2023} for the opportunity to explore and test methods discussed here in the context of human and fungal genetics analyses. 

\section*{Funding}

\noindent This work was partly supported by internal funding from the Universidad del Rosario (EMCS/PURE), ``Understanding regional genomic variation associated with risk and etiology of chronic aging diseases".

\section*{Declaration}

\noindent The author has no competing interests to declare.

\bibliographystyle{unsrt} 
\bibliography{biblio.bib}

\begin{thebibliography}{10}

\bibitem{JACC1-JEG-OKC}
J.E. Gallo, E.~Misas, J.G. McEwen, and O.K. Clay.
\newblock {{T}oward multiple {S}{N}{P} motif analyses of loci associated with
  phenotypic traits}.
\newblock {\em J. Am. Coll. Cardiol.}, 70:1539--1540, 2017.

\bibitem{GalloIJChy2020}
J.E. Gallo, J.E. Ochoa, H.R. Warren, E.~Misas, M.M. Correa, J.A.
  Gallo-Villegas, G.~Bedoya, D.~Aristiz\'abal, J.G. McEwen, M.J. Caulfield,
  G.~Parati, and O.K. Clay.
\newblock {{H}ypertension and the roles of the 9p21.3 risk locus: {C}lassic
  findings and new association data}.
\newblock {\em Int. J. Cardiol. Hypertens.}, 7:100050, 2020.

\bibitem{Dingsdag2021}
S.A. Dingsdag, O.K. Clay, and G.A. Quintero.
\newblock {{C}{O}{V}{I}{D}-19 severity, mi{R}-21 targets, and common human
  genetic variation. {L}etter regarding the article `{C}irculating
  cardiovascular micro{R}{N}{A}s in critically ill {C}{O}{V}{I}{D}-19
  patients'}.
\newblock {\em Eur. J. Heart Fail.}, 23:1986--1987, 2021.

\bibitem{BartonHaplotypes2023}
D.~Shipilina, A.~Pal, S.~Stankowski, Y.~F. Chan, and N.~H. Barton.
\newblock {{O}n the origin and structure of haplotype blocks}.
\newblock {\em Mol. Ecol.}, 32:1441--1457, 2023.

\bibitem{SabetiLander2024}
D.~Kotliar, S.~Raju, S.~Tabrizi, I.~Odia, A.~Goba, M.~Momoh, et~al.
\newblock {{G}enome-wide association study identifies human genetic variants
  associated with fatal outcome from {L}assa fever}.
\newblock {\em Nat Microbiol}, 9:751--762, 2024.

\bibitem{Uffelmann2021}
E.~Uffelmann, Q.Q. Huang, N.S. Munung, J.~de~Vries, Y.~Okada, et~al.
\newblock Genome-wide association studies.
\newblock {\em Nat. Rev. Methods Primers}, 1(59):1--21, 2021.

\bibitem{HollandBook1995GAs}
J.~H. Holland.
\newblock {\em Hidden Order: How Adaptation Builds Complexity}.
\newblock Addison-Wesley / Helix Books, Reading, MA, etc., 1995.

\bibitem{GoldbergBook1989GAs}
D.~E. Goldberg.
\newblock {\em Genetic Algorithms in Search, Optimization, and Machine
  Learning}.
\newblock Addison-Wesley, Reading, MA, etc., 1989.

\bibitem{EdwardsFoundations}
A.W.F. Edwards.
\newblock {\em Foundations of Mathematical Genetics}.
\newblock Cambridge University Press, Cambridge, etc., 2nd edition, 2000.

\bibitem{gephyrinYinYang}
S.~Climer, A.R. Templeton, and W.~Zhang.
\newblock {{H}uman gephyrin is encompassed within giant functional noncoding
  yin-yang sequences}.
\newblock {\em Nat. Commun.}, 6:6534, 2015.

\bibitem{Eichler2021}
P.~Ebert, P.~A. Audano, Q.~Zhu, B.~Rodriguez-Martin, D.~Porubsky, M.~J. Bonder,
  et~al.
\newblock {{H}aplotype-resolved diverse human genomes and integrated analysis
  of structural variation}.
\newblock {\em Science}, 372:1--13, 2021.

\bibitem{KGgenomes}
{1000 Genomes Project Consortium}.
\newblock A global reference for human genetic variation.
\newblock {\em Nature}, 526:68--74, 2015.

\bibitem{Beagle-v4-2013}
B.~L. Browning and S.~R. Browning.
\newblock {{I}mproving the accuracy and efficiency of identity-by-descent
  detection in population data}.
\newblock {\em Genetics}, 194:459--471, 2013.

\bibitem{DalyLanderHapblocks}
M.J. Daly, J.D. Rioux, S.F. Schaffner, T.J. Hudson, and E.~S. Lander.
\newblock High-resolution haplotype structure in the human genome.
\newblock {\em Nature Genet.}, 29:229--232, 2001.

\bibitem{WTCCC2007}
P.R. Burton, D.G. Clayton, L.R. Cardon, N.~Craddock, P.~Deloukas, A.~Duncanson,
  et~al.
\newblock Genome-wide association study of 14,000 cases of seven common
  diseases and 3,000 shared controls.
\newblock {\em Nature}, 447:661--678, 2007.

\bibitem{Ehret2016}
G.B. Ehret, T.~Ferreira, D.I. Chasman, A.U. Jackson, et~al.
\newblock The genetics of blood pressure regulation and its target organs from
  association studies in 342,415 individuals.
\newblock {\em Nat. Genet.}, 48:1171--1184, 2016.

\bibitem{VelezMM2023}
N.~V\'elez, N.~Vega-Vela, O.K. Clay, and C.M. Parra-Giraldo.
\newblock {{T}he structure of associations: {M}ethod insights from analyzing 28
  clinical isolates of \emph{{C}ryptococcus neoformans}}.
\newblock {\em Med. Mycol.}, 61:1--6, 2023.

\bibitem{MargalefBook}
R.~Margalef.
\newblock {\em Perspectives in Ecological Theory}.
\newblock University of Chicago Press, Chicago and London, 1968.

\end{thebibliography}

\end{document}